\documentclass[a4paper, twoside, superscriptaddress]{JHEP3}

\usepackage[section]{placeins}
\usepackage{bm}

\def\vp{{\varphi}}
\def\dvp{{\delta\varphi}}
\def\cs2{c_{\rm{s}}^2}

\newcommand\fnl{f_{\rm{NL}}}
\renewcommand\({\left(}
\renewcommand\){\right)}

\newcommand\be{\begin{equation}}
\newcommand\ee{\end{equation}}
\newcommand\bea{\begin{eqnarray}}
\newcommand\eea{\end{eqnarray}}
\newcommand\eq[1]{Eq.~(\ref{#1})}
\newcommand\eqs[2]{Eqs.~(\ref{#1}) and (\ref{#2})}

\def\r{{\rm{r}}}
\def\e{{\rm{e}}}
\newcommand{\bs}{{\varphi}}

\title{How the curvaton scenario, modulated reheating and an inhomogeneous end of inflation are related}
\author{Laila Alabidi \thanks{ LA is supported by the Science and Technologies
Facilities Council (STFC) under Grant PP/E001440/1.}\\Astronomy Unit, School of Mathematical Sciences, 
Queen Mary University of London, Mile End Road, E1 4NS, UK\\ E-mail: \email{l.alabidi@qmul.ac.uk}}
\author{Karim Malik\thanks{KAM is supported
in part by the Science and Technologies Facilities Council (STFC)
under Grant ST/G002150/1.}\\ Astronomy Unit, School of Mathematical Sciences, 
Queen Mary University of London, Mile End Road, E1 4NS, UK\\ E-mail: \email{k.malik@qmul.ac.uk}}
\author{Christian T. Byrnes\\Fakult\"at f\"ur Physik, Bielefeld University, 
Universit\"atsstrasse, D-33615 Bielefeld, Germany}
\author{Ki-Young Choi\thanks{ KYC is partly supported by the Korean
Government (KRF-2008-341-C00008) and by the second stage of Brain
Korea 21 Project in 2006.
}\\Department of Physics, Pusan National University, Busan, 609-735, Korea}
\abstract{
In this paper we analyse three models of the early universe, for which
the respective mechanisms for generating the curvature perturbation are considered
disparate. We find that in fact the mechanisms are very similar, and hence
explain why they give rise to a large non-gaussianity.
We show that the mechanism for generating the primordial
curvature perturbation, and hence the observable non-gaussianity, is
similar in both the Curvaton and Modulated Reheating models. In both
cases the model can be written in terms of an energy transfer between
the constituting fluids.  We then show that this is also true for the
mechanism of generating the curvature perturbation by symmetry
breaking the end of inflation. We then relate this to the non-gaussian
contribution to the curvature perturbation and find that it is
inversely proportional to the efficiency with which the curvature
perturbation is transferred between the fluids. For the first time, we generalise models
of modulated reheating to allow for a non-linear energy transfer rate.
}
\date{\today}
\preprint{BI-TP-2010-05, PNUTP-10-A04}
\begin{document}

\section{Introduction}

One of the greatest successes of inflationary cosmology is to
provide a mechanism to generate the spectrum for the
primordial density perturbations, in excellent agreement with recent
observations of the Cosmic Microwave Background (CMB)
\cite{Komatsu:2010fb} and galaxy surveys
(e.g.~\cite{Tegmark:2006az}). In the most popular single field
inflation models, the same field that drives inflation is also
responsible for the generation of the spectrum during
inflation. However, in recent years models in which these tasks are
 separated have become prominent, in particular the curvaton
scenario \cite{DLDW,MT,ES}, modulated reheating \cite{Dvali,Kofman:2003nx,Bernardeau:2004zz,Zal}, and
an inhomogeneous end of inflation \cite{bern1,bern2,AL-eoi,Sasaki:2008uc,Naruko:2008sq} (here the separation is
incomplete).  Although these models were usually considered to be very
different, taking the mechanism that generates the primordial spectrum
of perturbations to categorise the models we show that these models
are in fact very similar, as the generation mechanism of the
perturbations is the same. This also explains why all three models can
give rise to large non-gaussianity.

The same observations cited above also 
show that the spectrum of primordial density perturbations is
predominantly gaussian in nature with a possible minor deviation
\cite{Smith:2009jr,Rudjord:2009mh}. Single field canonical models of
inflation are known to give a non-gaussian contribution
which is slow roll suppressed \cite{Maldacena:2002vr}; 
this level of non-gaussianity is undetectable by current
technology.  
It has been shown that if the field that generated the primordial
curvature perturbation interacts with another fluid, such as in the
curvaton \cite{Lyth:2002my,ML} or modulated reheating
\cite{Dvali,Kofman:2003nx,Bernardeau:2004zz,Zal}, or if inflation is driven by more than one field,
see e.g.~\cite{Alabidi:2006hg,Byrnes:2008wi,Byrnes:2009qy}, 
then the non-gaussianity can be enhanced to a detectable level.\\

In this paper we evaluate the curvature perturbation $\zeta$ and the
non-linearity parameter $\fnl$ for models of modulated reheating, the
curvaton model, and the inhomogeneous end of inflation. We
investigate the relationship between these models and highlight the
underlying mechanism of generation and the conditions for enhancing
$\fnl$.
Although at first glance these models appear to be very disparate, we
show that the generation mechanism in all three cases is similar: all
three models can be written in terms of an energy transfer between the
constituting fluids; the energy transfer itself is controlled by a light
scalar field; the fluctuations of the scalar field are then
``inherited'' after its decay by the curvature perturbation.  
In the curvaton and modulated reheating models, the curvature
perturbation can be parametrised by the efficiency with which the
curvature perturbation is translated from the field that generates the
primordial curvature perturbation to the radiation field and hence is
the determining factor in $\fnl$. The same is true for the
inhomogeneous end of inflation case, however the efficiency parameter
here refers to the efficiency by which the curvature perturbation is
translated from one scalar field to the one that generates the
primordial density perturbation.

For the first time we consider an energy transfer which depends 
non-linearly on the energy density in the modulated reheating scenario.
We show how significantly more general scenarios lead to a curvature
perturbation with the same functional form but modified coefficients,
which affects the value of $\fnl$ but it still becomes large in the same
limit as in the standard case of a linear transfer. Unlike previous
calculations, we do not resort to solving the
equations of modulated reheating
numerically, which makes the derivation more transparent.

This paper is organised as follows: in Section (\ref{ET}) we present
the equations governing the energy transfer between fluids, in Section
(\ref{sec-radiation}) we evaluate the curvature perturbation for
models in which a scalar field decays into radiation, in Section
(\ref{eoi}) we calculate the curvature perturbation using a different
approach to that derived in previous work for the inhomogeneous end of
inflation. We present the results for the non-gaussianity parameter
$\fnl$ in Section (\ref{fnl}) and discuss them in Section
(\ref{disc}). We use the convention of decomposing the curvature
perturbation in terms of the first order $\zeta_1$ and second order
$\zeta_2$ contributions as follows
\be
\zeta=\zeta_1+\frac{1}{2}\zeta_2\, ,
\ee
where $\zeta_1$ is the linear (gaussian) perturbation and $\zeta_2$ is
a gaussian squared, see e.g.~Ref.~\cite{MW2008}.

\section{Energy Transfer}
\label{ET}
The line element for a Friedmann-Robertson-Walker spacetime including
scalar perturbations, without yet specifying a particular gauge, is
\be
\label{ds2}
ds^2=-(1+2\phi)dt^2+2aB_{,i}dt dx^i
+a^2\left[(1-2\psi)\delta_{ij}+2E_{,ij}\right]dx^idx^j \,, 
\ee
where we use the notation and conventions of Refs.~\cite{MW2008} with
the curvature perturbation, $\psi$, the lapse function, $\phi$, and
scalar shear, $a^2\dot E - aB$, and where $\delta_{ij}$ denotes the
flat background metric and $X_{,i}\equiv\partial{}X/\partial{}x^i$.

Allowing for the exchange of energy between fluids, the equations
governing the evolution of the density perturbation are given by
Refs.~\cite{KS,MW,MW2008}. From the local energy-momentum equation,
\be
\nabla_\mu{}T^{\mu\nu}_{(\alpha)}=Q^\nu_{(\alpha)}\,,
\ee
we have the evolution equation for the energy density of a particular
fluid in the background
\be\label{rho-evolve}
\dot\rho_\alpha+3H(\rho_\alpha+P_\alpha)=Q_\alpha\,,
\ee
and at first order in the perturbations on superhorizon scales
\be\label{delta-rho-evolve}
\dot{\delta\rho}_\alpha+3H(\delta\rho_\alpha+\delta{}P_\alpha)
-3(\rho_\alpha+P_\alpha)\dot{\psi}+Q_\alpha\psi-\delta{}Q_\alpha=0\,,
\ee
where $\rho$ is energy density, $P$ is pressure, subscript $\alpha$
denotes the fluid species, $Q_\alpha$ is the energy transfer to the
$\alpha$ fluid and $T^{\mu\nu}$ is the stress-energy tensor.

Note that in Sections (\ref{linear}) to (\ref{hybrid}) we use the
longitudinal gauge, $a^2\dot{E}-aB=0$, and we ignore anisotropic stress so $\phi=\psi$, whereas
in Sections (\ref{curvaton}) and (\ref{eoi}) we use the flat gauge,
$\psi=0$, for convenience.

\section{Models in which a scalar field decays into radiation}
\label{sec-radiation}
We now consider three different forms for the energy transfer: in
model (\ref{linear}) $Q_m$ is a linear function of $\rho_m$, in model
(\ref{higher}) it is a higher order function of $\rho_m$ and in
model (\ref{hybrid}) it is a function of a combination of powers. 
These three models are all examples of modulated reheating, 
in which the decay rate from the inflaton to radiation is modulated 
by a subdominant scalar. Since the Univere is matter dominated 
before decay and radiation dominated after decay, changes in the 
decay time lead to changes in the expansion and hence effect the 
curvature perturbation \cite{Dvali}.
In
section (\ref{curvaton}) we present the generic results for the
curvaton scenario, in which the rate of decay is homogeneous.

\subsection{$Q_m$ as a linear function of $\rho_m$}
\label{linear}
We model the energy transfer $Q_\alpha$ according to the standard
choice 
\be
Q_\alpha=-\Gamma\rho_\alpha\,,
\ee
where $\Gamma$ is the decay rate of the $\alpha$ field into radiation.
In this section we consider modulated reheating in
which the decay rate of a
light scalar field $\vp$ 
varies, that is
$\Gamma=\Gamma(\bs)$,  
and in Section~(\ref{curvaton})
we analyse the curvaton ($\sigma$) model
in which $\Gamma$ is taken to be constant. For both these models 
the background evolution equations for an oscillating scalar field 
(i.e.~matter type) fluid and radiation are \cite{MW,Dvali}
\bea
\label{BG}
\dot\rho_m=-3H\rho_m-\Gamma\rho_m\,,\nonumber\\
\dot\rho_r=-4H\rho_r+\Gamma\rho_m\,,
\eea
and we assume that the decay is into Standard Model
radiation.
Evaluating \eq{rho-evolve} for the oscillating scalar field,
noting that
$P_m=\delta{}P_m=0$
and $\delta{}Q_m=-\delta\Gamma\rho_m-\Gamma\delta\rho_m$, 
which follows from (\ref{BG}),
we get
\be
\frac{\dot\delta\rho_m}{\rho_m}
+3H\frac{\delta\rho_m}{\rho_m}-3\dot\phi
+\Gamma\phi+\Gamma\frac{\delta\rho_m}{\rho_m}+\delta\Gamma=0\,.
\ee
Introducing the density contrast
$\delta_\alpha=\delta\rho_\alpha/\rho_\alpha$ to rewrite the above
equation in a more compact form, and using \eq{BG} we get
$\dot\delta\rho_m/\rho_m=\dot\delta_m-\delta_m(3H+\Gamma)$ and
\be
\dot\delta_m=3\dot\phi-\Gamma(\phi+\delta_\Gamma)\,,
\ee
where $\delta_\Gamma=\delta\Gamma/\Gamma$.  Similarly, for the
radiation fluid, where $P_\r=\rho_\r/3$ and $Q_\r=\Gamma\rho_m$, we
find
\be
\dot\delta_r=4\dot\phi
+\Gamma\frac{\rho_m}{\rho_\r}(\phi+\delta_\Gamma+\delta_m-\delta_\r)
\,.
\ee

Finally we have from the time component of the Einstein equations
$G_{\mu\nu}=8\pi{}GT_{\mu\nu}$  on large scales \cite{MW}
\be
H\dot\phi+H^2\phi=-\frac{4\pi{}G}{3}(\rho_m\delta_m+\rho_\r\delta_\r)\,.
\ee

In Ref.~\cite{Dvali} Modulated Reheating is studied in the so 
called `purely forced' case, where the metric perturbations
are sourced only by the perturbations in the rate of decay. To
derive the curvature perturbation for this case we assume that the
perturbations in the matter and radiation fields are subdominant,
hence leaving us with two equations
\bea
H\dot\phi_{\rm{mod}}+H^2\phi_{\rm{mod}}=0\nonumber\,, \\
3\dot\phi_{\rm{mod}}=\Gamma(\phi_{\rm{mod}}+\delta_\Gamma)\,,
\eea
where the subscript ``$mod$'' refers to `modulated reheating'. The
above equations then give
\be
\label{Phi-Beta}
\phi_{\rm{mod}}=-\frac{\Gamma}{\Gamma+3H}\delta_\Gamma
=-\beta\frac{\delta\Gamma}{\Gamma}\,.
\ee
Here $\beta$ is proportional to the decay rate with $\beta\simeq1$
corresponding to instant reheating.

The curvature perturbation on uniform density hypersurfaces is defined as
\be
\label{def_zeta}
\zeta_{\rm{mod}}=-\phi_{\rm{mod}}-H\frac{\delta\rho}{\dot\rho}\,.
\ee
where $\delta\rho/\dot{\rho}=\sum_i\delta\rho_i/\sum_j\dot{\rho_j}$. Evaluating the RHS of \eq{def_zeta} in the ``forced case'' (i.e.~when
$\delta\rho/\rho\ll \phi$) gives 
\be
\label{def_zeta2}
\zeta_{\rm{mod}}=-\phi_{\rm{mod}}\,,
\ee
and finally, substituting \eq{Phi-Beta} into \eq{def_zeta2} we get
\be\label{zeta_mod_final}
\zeta_{\rm{mod}}=\beta\frac{\delta\Gamma}{\Gamma}\,.
\ee

Assuming now that $\Gamma$ depends only on one light scalar field, $\bs$,
(Ref.~\cite{Zal} considers a more general case) a simple Taylor
expansion results in
\be
\Gamma=\Gamma_{0}+\frac{\partial\Gamma}{\partial\bs_*}\delta\bs_*
+\frac{1}{2}\frac{\partial^2\Gamma}{\partial\bs_*^2}\delta\bs_*^2\,,
\ee
where the subscript $*$ refers to horizon exit, this gives the curvature perturbation as
\be
\label{zeta-mod}
\zeta_{\rm{mod}}=\frac{\beta}{\Gamma}\(\Gamma_{\bs_*}\delta\bs_*
+\frac{\Gamma_{\bs_*\bs_*}}{2}\delta\bs_*^2\)\,,
\ee
where $\Gamma_{\bs_*}$ and $\Gamma_{\bs_*\bs_*}$ are the first and second
derivatives with respect to $\bs_*$.

In sections (\ref{curvaton}) and (\ref{eoi}) we define the efficiency
as the ratio of the first order curvature perturbation of a field
prior to its decay with respect to its value post decay, and hence we
construct similar parameters for this model for completeness. The field
$\bs_*$ can be related to a curvature perturbation of
$\zeta_\bs=(\ddot{\bs_*}+V_{,\bs_*})\delta\bs_*/(3\dot{\bs_*}^2)\equiv b\delta\bs_*$,
see e.g.~Ref.~\cite{balloon}, then the efficiency parameter is
\be\label{c1}
c_1=\frac{\zeta_{\rm{mod}~1}}{\zeta_\bs}
=\frac{\beta\Gamma_{,\bs_*}}{b\Gamma}\,,
\ee
\be\label{c2}
c_2=\frac{\zeta_{\rm{mod}~2}}{\zeta_\bs^2}
=\frac{\beta\Gamma_{,\bs_*\bs_*}}{b^2\Gamma}\,.
\ee
The overall curvature perturbation for this model
can now be written as
\be
\label{zeta-mod-new}
\zeta_{\rm{mod}}=bc_1\delta\bs_*+b^2c_2\delta\bs_*^2\,.
\ee

\subsection{$Q_m$ as a higher order function of $\rho_m$}
\label{higher}
In this case we define the energy transfer as a function of a higher
power of the energy density, similar to the model recently considered
in the context of dark energy decay in \cite{maartens},
\be
Q_m=\Gamma\rho_m^n\,.
\ee
In order to avoid singular behaviour in the limit of small $\rho_m$ we require $n>0$.

Following the same steps as in section (\ref{linear}), i.e.~only
considering $\delta\rho=0$, we find that the curvature perturbation on
superhorizon scales has the same form as \eq{zeta-mod}, with
$\beta$ now also a function of the energy density:
\be\label{beta-non-linear}
\beta=\frac{\Gamma\rho_m^{n-1}}{3H+\Gamma\rho_m^{n-1}}\,.  
\ee
In order for $\rho_m$ to decay then, the decay rate must exceed the
expansion rate of the Universe, $\rho_m\ll10^{-12}$ in Planck Units,
and since $H$ decays as $\rho_m^{1/2}$, then in order for
$H/\Gamma\rho_m^{n-1}<1$ we require $n<3/2$.

\subsection{$Q_m$ as a function of a combination of 
powers of $\rho_m$}
\label{hybrid}
In this case we define the energy transfer function as
\be
Q_m=\Gamma\(\mathcal{A}\rho_m^p+\mathcal{B}\rho_m^q\)\,,
\ee
where $0<p<q$. Then $\zeta$ has the same functional form as \eq{zeta-mod}
but with $\beta$ redefined as
\be
\beta=\frac{\Gamma\(\mathcal{A}\rho_m^{p-1}
+\mathcal{B}\rho_m^{q-1}\)}{3H+\Gamma\(\mathcal{A}\rho_m^{p-1}
+\mathcal{B}\rho_m^{q-1}\)}\,.
\ee

\subsection{The curvaton scenario: $\delta\Gamma=0$}
\label{curvaton}
We now consider the second scenario, where $\delta\Gamma=0$. This is
the case in the curvaton model. The energy density of the curvaton is 
highly subdominant compared to the inflaton field during inflation, 
but afterwards, while it oscillates about the (quadratic) minimum of 
its potential, its energy density decays like matter. This energy 
component grows relative to the radiation density generated
 by the inflaton decay products until the curvaton also decays into 
 radiation \cite{DLDW,MT,ES}. In this scenario the curvature perturbation 
 is solely generated from perturbations in the curvaton field.
Using a quadratic potential for the curvaton, the curvature perturbation in this model was found to be
\be
\label{zeta-curv}
\zeta_{\rm{curv}}=\frac{2r_1}{3}\(\frac{\delta\sigma_{1*}}{\sigma_{0*}}
+\frac{1}{2}\(\frac{\delta\sigma_{1*}}{\sigma_{0*}}\)^2\)
\ee
where $\sigma$ has been split as $\sigma=\sigma_{0*}+\delta\sigma_{1*}$, and
$r_1$ parametrises the contribution of the $\sigma$ field to the
overall curvature perturbation post $\sigma$ decay
\be\label{r1}
r_1=\frac{\zeta_{\rm{curv}~1,\rm{post}~\sigma~\rm{decay}}}
{\zeta_{\rm{curv}~1,\rm{pre}~\sigma~\rm{decay}}}\,.
\ee
Using second order perturbation theory $\zeta_{\rm{curv}}$ is given by \cite{ML}
\be
\zeta_{\rm{curv}}=\frac{2}{3}r_1\frac{\delta\sigma_{1*}}{\sigma_{0*}}
+\frac{1}{3}\left(r_1+\frac{2}{3}r_2\right)\(\frac{\delta\sigma_{1*}}{\sigma_{0*}}\)^2
\,,
\ee
where
\be
r_2=\frac{\zeta_{\rm{curv}~2,\rm{post}~\sigma~\rm{decay}}}
{\zeta_{\rm{curv}~1,\rm{pre}~\sigma~\rm{decay}}^2}\,.
\ee

\section{Inhomogeneous End of Inflation}
\label{eoi}
In previous work \cite{bern1,bern2,Lyth:2005qk}, 
it has been shown that the curvature perturbation 
can be generated at the end of inflation if there is 
an ultra light scalar field sub-dominant to the inflato 
during inflation. This second field does not play a 
role in the inflationary dynamics, but serves to perturb 
the inflaton trajectory from a straight line, potentially 
resulting in the generation of non-gaussianity. 
This concept was generalised to a two-field hybrid 
model of inflation in \cite{AL-eoi} and was found to 
give a measurable $\fnl$. The potential describing this model is given by:
\be
\label{inhom-pot}
V=V_0-\frac{m^2}{2}(\vp^2+\sigma^2)
-(f\vp^2+g\vp\sigma+h\sigma^2)\frac{\chi^2}{2}+\frac{m_\chi^2}{2}\chi^2\,,
\ee
where $\chi$ is the waterfall field which is held at zero during
inflation, and $f,g$ and $h$ define the coupling of the inflaton
fields to the waterfall field. In this case $f\neq{}g\neq{}h$ and
hence the surface at the end of inflation is best defined by an
ellipse.  
For simplicity we assume an instantaneous decay of
  the scalar fields into radiation at the end of inflation, for a more
  detailed discussion of this point see
  Refs.~\cite{Lyth:2005qk,Sasaki:2008uc}. 

In contrast to previous work carried out on this and similar models
\cite{AL-eoi,bern1,bern2,Sasaki:2008uc,Naruko:2008sq} which used the $\delta{}N$ formalism, 
we use the perturbative approach in calculating $\fnl$. 
Starting with the definition of the
curvature perturbation at horizon exit,
\be
\zeta_\e=-H\frac{\delta\rho_\e}{\dot{\rho_\e}}\,,
\ee
where the subscript ``$\e$'' refers to the end of inflation.  We take the
$\sigma=0$ trajectory and split $\vp_\e$ as $\vp_\e=\vp_{0\e}+\dvp_\e$,
and thus,
\be
\label{zeta_ee}
\zeta_\e=\frac{1}{2\eta_s}\(2\frac{\dvp_\e}{\vp_{0\e}}
+\(\frac{\dvp_\e}{\vp_{0\e}}\)^2\)\,.
\ee

In order to evaluate the spectrum and bi-spectrum for this model we
will need to calculate terms of the form
$<\dvp_\e(\mathbf{k_1})\dvp_\e(\mathbf{k_2})>$, but since we only know
their values at horizon exit we re-write $\dvp_\e$ in terms of
$\dvp_*$:
\be
\dvp_\e=\dvp_\e(\vp_*+\dvp_*,\sigma_*+\delta\sigma_*)
=\frac{\partial\vp_\e}{\partial\vp_*}\dvp_*
+\vp_\e'\delta\sigma_*+\frac{\vp_\e''}{2}\delta\sigma_*^2 
\ee
where $\vp_\e'=\partial\vp_\e/\partial\sigma=-g/2f$,
$\vp_\e^{''}=-(2h-g^2/(2f))/(2f\vp_\e)$,
$\partial\vp_\e/\partial\vp_*=e^{\eta_sN}$ \cite{AL-eoi}, $\eta_s$ is
the second derivative slow roll parameter and $N$ is the logarithmic ratio
of the scale factor at the end of inflation with respect to its value
when scales of cosmological interest exited the horizon\cite{book2}.

Assuming that the curvature perturbation is generated predominantly
from the $\sigma$ field, then Eq.~(\ref{zeta_ee}) becomes
\be
\zeta_{\sigma}=\frac{\vp_\e'}{\eta_s\vp_\e}\delta\sigma_*
+\frac{1}{2\eta_s\vp_\e^2}\(\vp_\e^{'2}+\vp_\e\vp_\e^{''}\)\delta\sigma_*^2\,,
\ee
and since we are mainly interested in highlighting the underlying
physical mechanism, we have used a simplifying \cite{Lyth:2002my}
calculation to get the second order contribution, which explains why
our results do not match those of Ref.~\cite{AL-eoi} at second order
exactly.

We now define the parameters
\be
b_1=\frac{\zeta_{\sigma~1}}{\zeta_{\vp~1}}=\vp_\e'e^{-\eta_sN}\,,
\ee
and
\be
b_2=\frac{\zeta_{\sigma~2}}{\zeta_{\vp~1}^2}
=\frac{\eta_s}{2}e^{-2\eta_sN}\(\vp_\e'^2+\vp_\e\vp_\e^{''}\)\,,
\ee
in analogy with the parameters $c_1$ and $c_2$ above.
Therefore the curvature perturbations generated by the $\sigma$ field
can be written as
\be
\zeta_\sigma=\frac{1}{\eta_s}\(b_1\(\frac{\delta\sigma_*}{\vp_*}\)
+\frac{b_2}{\eta_s}\(\frac{\delta\sigma_*}{\vp_*}\)^2\)\,.
\ee

This model can be recast in the form of two interacting fluids
by writing \eqs{rho-evolve}{delta-rho-evolve} in the slow roll approximation for this
case
\bea
\dot{\rho}_\vp&=&-Q_\sigma=-V_{,\sigma}\dot{\sigma}=0\,,\nonumber\\
\dot{\delta\rho}_\vp&=&-\delta{}Q_\sigma
=\frac{V_{,\sigma\sigma}^2}{3H}\delta\sigma_*^2\,,
\eea
showing more clearly the relation to the previous models and hence
explaining the similar results in terms of the non-linearity
parameter, as detailed in the next section. Note that we are only
modelling the energy transfer between fields and not the decay
of the field into radiation. We assume an instantaneous decay into
radiation at the end of inflation, for a discussion of this point see 
Refs.~\cite{Lyth:2005qk,Sasaki:2008uc}. 
The energy of the $\vp$ field is transferred to the radiation at the
end of inflation and the perturbation of this energy transfer is
dominantly controlled by the perturbation of the $\sigma$ field. After
the decay of the $\vp$ field, the energy of the radiation was endowed
from the $\vp$ field but the perturbation is inherited from the
$\sigma$ field.

\section{The Non-Gaussianity Parameter $\fnl$}
\label{fnl}
From \eq{zeta-mod-new} we find that the non-linearity parameter $\fnl$ for
the modulated reheating case is given by 
\be
\fnl=\frac{5}{6}\frac{c_2}{c_1^2}\,,
\ee
which in terms of $\beta$ is
\be
\label{fnl-mod}
\fnl=\frac{5}{12}\frac{\Gamma}{\beta}\frac{\Gamma_{\bs_*\bs_*}}{\Gamma_{\bs_*}^2}\,.
\ee 
Using the functional form $\Gamma=\Gamma_1(1+\delta\bs_*/\bs_*)^2$, following \cite{Zal},
Eq.~(\ref{fnl-mod}) gives
\be\label{fnlm1}
\fnl=\frac{5}{24\beta}\,,
\ee
and $\beta$ depends on the decay rate. This inverse dependence on
$\beta$ is also valid for the new models discussed in Sections (\ref{higher}) and (\ref{hybrid}).

The non-linearity parameter $\fnl$ for the curvaton model is derived
in the simplest case from \eq{zeta-curv} to be \cite{Lyth:2002my}
\be
\label{fnl-curv1}
\fnl=\frac{5}{4r_1}\,,
\ee
and using second order perturbation theory \cite{ML}
\be
\label{fnl-curv2}
\fnl=\frac{5}{4r_1}+\frac{5r_2}{6r_1^2}-\frac{5}{3}\,.
\ee
Eqs.~(\ref{fnl-curv1}) and (\ref{fnl-curv2}) were evaluated for the
potential $V(\sigma)=m^2\sigma^2/2$.

Finally we have for the inhomogeneous end of inflation:
\be
\fnl=\frac{5}{6}\frac{b_2}{b_1^2}\,.
\ee

\section{Discussion}
\label{disc}
In this paper we have studied models of the early universe dominated
by a scalar field that then decays into radiation, and a model in
which two scalar fields drive inflation where the curvature
perturbation is generated primarily at the end of inflation. Using the
energy transfer $Q=\Gamma\rho$ to characterise the models, the former
case can be further split into models with a homogeneous $\Gamma$ and
varying $\rho$, Curvaton models, and models with a varying decay rate
$\Gamma$ and homogeneous energy density $\rho$, Modulated reheating.

We then evaluated $\fnl$ for the simplest case of a homogeneous
$\Gamma$,  and for various functional
forms of the energy transfer parameter $Q_\vp$.  
We found that in the varying $\Gamma$ case $\fnl$ is inversely
proportional to the decay efficiency $\beta$,
even in the new case we consider of a non-linear energy transfer, $Q=\Gamma\rho^n$
or $Q=\Gamma\(\mathcal{A}\rho_m^p+\mathcal{B}\rho_m^q\)$, which parametrises the
rate at which the scalar field decays into radiation, as well as the
energy density of the scalar field prior to its' decay.  Reparametrising
our equations shows that $\fnl$ is inversely proportional
to $c_1^2$, where $c_1$ is the ratio of $\zeta$ prior to the decay of 
the light scalar $\vp$ to it's value post $\vp$ decay. Similarly in the case
where the curvature perturbation is generated by the inhomogeneous
end of inflation, we found that $\fnl$ is inversely proportional to
the square of the parameter $b_1$,  the ratio of the first
order curvature perturbation sourced by $\sigma$ to the curvature
perturbation sourced by $\vp$ at horizon exit. Previous work \cite{ML}
(which we reproduced here) already showed a similar result for the
curvaton model; $\fnl$ is inversely proportional to $r_1$, the
ratio of the first order curvature perturbation after the decay of
$\sigma$ with respect to the value of the curvature perturbation 
prior to its decay, \eq{r1}.

Our results show that the mechanism of generating the curvature
perturbation in these models is similar, and can be modelled as energy
transfer between interacting fluids: the fluctuations are generated in
the first fluid, a massless scalar field, and are then inherited by
the second one into which it decays. This therefore explains why the
condition for enhancing $\fnl$ is similar; it is the efficiency with
which the energy density or curvature perturbation is translated
between the fluids that determines the level of non-gaussianity.
 
Although the three models we have discussed are physically very 
different and neither share the same fields nor generate the curvature
perturbation during the same epoch, in all cases the curvature perturbation
is related to an energy transfer. This energy transfer is from a light scalar
field into some other fluid into which it (and any other fields present) decay,
see Table(\ref{table}). We have explicitly shown that in all three cases it
is the efficiency with which the energy density or curvature perturbation
is translated between the fluids that determines the level of non-gaussianity.

\begin{table}

\begin{tabular}{|c|c|c|c|c|}
\hline
&\multicolumn{2}{|c|}{Energy Transfer}&Modulating Field&Background\\
Model& from & to & ($\delta Q$)&energy
from\\
\hline
MR& dominant &radiation & a light scalar field
& dominant \\
&oscillating scalar &&$(\propto\delta\vp$)&oscillating scalar\\
\hline
Curvaton& curvaton & radiation & curvaton 
& radiation \\
&&&($\propto\delta \sigma$)&before curvaton decay\\
\hline
IEI& dominant & waterfall &
subdominant & dominant \\
&inflaton ($\varphi$) &field ($\chi$)&inflaton ($\propto\delta\sigma$)&inflaton ($\varphi$)\\
\hline
\end{tabular}
\caption{We summarise the generation of large non-Gaussianity in the three models analysed in this paper. The 
second and third column show the transfer of energy
and the fourth column shows the source fluid 
which modulates the energy transfer and generates the
curvature perturbation. The fifth column shows the source of the dominant background
energy density. Large non-Gaussianity is only possible when the source
of the background energy and that of the perturbations are different.}
\label{table}
\end{table}

\section{Acknowledgements}
We would like to thank Adam Christopherson and James Lidsey for useful
discussions. 

\bibliographystyle{JHEP}
\bibliography{modulated-reheating}
\end{document}